\documentclass[aps,pre,preprint,floats]{revtex4}
\input{epsf}
\input{psfig.sty}
\usepackage{amssymb}
\usepackage{graphicx}
\begin{document}
\def\ra{{r_{\rm Hb}^c}}
\def\rb{{r_{\rm Hb}^*}}
\def\rc{{r_{\rm Hb}^{sol}}}
\def\rx{{r_{\rm Hb}^x}}
\def\rhb{r_{\rm Hb}}
\def\s{{\frac{1}{2}}}
\def\x{{x_{\rm Hb}}}
\def\Ra{{r_{\rm sc}^c}}
\def\Rb{{r_{\rm sc}^*}}
\def\Rc{{r_{\rm sc}^{sol}}}
\def\Rx{{r_{\rm sc}^x}}
\def\X{{x_{\rm chiral}}}
\def\dt{{\Delta t}}
\def\go{{G$\rm\bar{o}$}}
\def\LJf{{LJ$^{fix}$}}
\def\LJa{{LJ$^{ang}$}}
\def\Solf{{Sol$^{fix}$}}
\def\Sola{{Sol$^{ang}$}}
\def\Qi{{Q_{int}}}
\def\qi{{\frac{Q_{int}}{N}}}
\def\q{{\langle\bar{Q}\rangle}}
\def\zqq{{Z_Q\left[\langle\bar{Q}\rangle\right]}}
\def\pdd{{p_{dd}}}
\def\cdd{{[C_{dd}]}}
\def\ps{{p_{s}}}
\def\cs{{[C_{s}]}}
\def\NHOC{{NH$\cdots$OC }}
\def\vHb{{v_{\rm Hb}}}
\def\fsq{{f_{i,j}^{\text{\em sq}}}}
\def\psq{{P_{\text{\em sq}}}}
\def\HVs{{\rm{H}_{V_s}}}
\def\HDS{{\rm{H}_{\Delta S}}}
\def\ra{{r_{\rm Hb}^c}}
\def\rb{{r_{\rm Hb}^*}}
\def\rc{{r_{\rm Hb}^{\text{\em sol}}}}
\def\rx{{r_{\rm Hb}^x}}
\def\rhb{r_{\rm Hb}}
\def\s{{\frac{1}{2}}}
\def\x{{x_{\rm Hb}}}
\def\Ra{{r_{\rm sc}^c}}
\def\Rb{{r_{\rm sc}^*}}
\def\Rc{{r_{\rm sc}^{\text{\em sol}}}}
\def\Rx{{r_{\rm sc}^x}}
\def\X{{x_{\rm chiral}}}
\def\dt{{\Delta t}}
\def\go{{G$\rm\bar{o}$}}
\def\LJf{{LJ$^{fix}$}}
\def\LJa{{LJ$^{ang}$}}
\def\Solf{{Sol$^{fix}$}}
\def\Sola{{Sol$^{ang}$}}
\def\Qi{{\bar{Q}_{\text{\em int}}}}
\def\qi{{\frac{Q_{\text{\em int}}}{N}}}
\def\Q{{\bar{Q}}}
\def\Qx{{\bar{Q'}}}
\def\Qij{{{Q'}_{ij}}}
\def\Qijk{{{Q'}_{ij,(k)}}}
\def\q{{\langle\bar{Q}\rangle}}
\def\HHb{{\rm H_{{\rm Hb}}}}
\def\Hco{{{\rm H}_{\text{\em coil}}}}
\def\UHb{{{\rm U_{Hb}}(r,\x)}}
\def\Dij{{\Delta_{i,j}}}
\def\Dijs{{{\Delta}_{ij,(s)}}}
\def\Dijt{{{\Delta}_{ij,(t)}}}
\def\D{{\{\Delta_{i,j}\}}}
\def\Z{{Z\left[\{\Delta_{i,j}\}\right]}}
\def\zqq{{Z_{\bar{Q}}\left[\langle\bar{Q}\rangle\right]}}
\def\zq{{Z_{\bar{Q}}}}
\def\M{{M^*}}
\def\sx{{\sigma_{\frac{1}{L}}}}
\def\pdd{{p_{dd}}}
\def\cdd{{[C_{dd}]}}
\def\ps{{p_{s}}}
\def\cs{{[C_{s}]}}
\def\NHOC{{NH$\cdots$OC }}
\def\vHb{{v_{\rm Hb}}}
\def\fsq{{f_{i,j}^{\text{\em sq}}}}
\def\psq{{P_{\text{\em sq}}}}
\def\HVs{{\rm{H}_{V_s}}}
\def\HDS{{\rm{H}_{\Delta S}}}

\title{Non-native $\beta$-sheet formation: insights into
protein amyloidosis}

\author{Chinlin Guo$^{\dag\ddag}$, Herbert Levine$^\ddag$, 
and David A. Kessler$^*$}
\affiliation{ $^\dag$Department of Molecular Cell Biology, 
Harvard University,
16 Divinity Avenue, Room 3007, Cambridge, MA 02138} 
\affiliation{ $^\ddag$Department of Physics, University of
California, San Diego
9500 Gilman Drive, La Jolla, CA 92093-0319} 
\affiliation{ $^*$Department of Physics, Bar-Ilan
University, Ramat-Gan, Israel}

%\maketitle

\begin{abstract}

Protein amyloidosis is a cytopathological process 
characterized by the formation of highly $\beta$-sheet-rich 
fibrils. How this process occurs and how to prevent/treat
the associated diseases are not completely understood. Here, we
carry out a theoretical investigation of sequence-independent
$\beta$-sheet formation, based on recent findings regarding the
cooperativity of hydrogen-bond network formation. 
Our results strongly suggest that {\it in vivo} $\beta$-sheet 
aggregation is induced by inter-sheet stacking dynamics. This 
leads to a prediction for the minimal length of susceptible 
polymer needed to form such an aggregate. Remarkably, the prediction 
corresponds quite well with the critical lengths detected in 
poly-glutamine-related diseases. Our work therefore provides a 
theoretical framework capable of understanding the underlying 
mechanism and shedding light on therapy strategies of 
protein amyloidosis.

\end{abstract}
\maketitle

\bigskip

\noindent Protein amyloidosis \cite{Dill} is seen in the prion-related
transmissible spongiform encephalopathies such as 
Creutzfeldt-Jakob disease \cite{TSE} and nine fatal 
poly-glutamine (poly-Q) related neurodegenerative disorders
including Huntington disease, Kennedy disease, six 
spinocerebellar ataxias, and dentatorubral pallidoluysian 
atrophy \cite{polyQ}. It has also been found that normal 
peptides can undergo misfolding to form amyloid aggregate 
\cite{Dobson}. Thus, protein amyloidosis can generally involve a 
sequence-independent mechanism; the critical questions then
would be how this process is initiated and how to prevent its
onset.

Although many different proteins can form amyloidosis \cite{Dobson}, 
it has been shown that the agents in aggregation-prone peptides
are their N-terminal oligopeptide repeats, which have a tendency
to form non-native $\beta$-sheets that in turn can serve as 
templates to aggregate additional mis/un-folded peptides into 
elongated fibrils \cite{polyQ,oligo-prion}.
The rate-limiting step in this process is the nucleation of the
non-native $\beta$-sheets \cite{nucleation}; moreover, the 
aggregation tendency correlates well with the length of repeats 
$\Delta l$ \cite{polyQ,oligo-prion}. For instance, in the Huntington 
disease gene, the ``normal'' range of poly-Q repeats is 6 to 34 CAGs, 
and the pathological forms involve tracts of 37 or longer repeats 
\cite{polyQ}. Likewise, a deletion of prion N-terminal repeats can 
suppress its spontaneous aggregation rate \cite{oligo-prion}.

What is the biological implication of these facts?
In principle, a longer $\Delta l$ implies a larger non-native $\beta$-sheet
formation. Under normal physiological conditions, the nucleation of 
non-native $\beta$-sheets is suppressed by intracellular scavengers, 
chaperones and the ubiquitin-proteosome system which detect and 
convert/degrade misfolded peptide intermediates \cite{chaperone,protease}. 
It has however been found that when $\Delta l$ exceeds the pathological 
threshold, scavengers do not work; instead, they become non-functionally 
engaged with the non-native structure \cite{UPS}. This impairment of 
molecular scavengers indicates that the formation of large non-native 
$\beta$-sheets can undergo a rapid ``two-state-like'' transition, i.e., 
proceeding without any intermediate of long enough detectable lifetime. 
It also implies that the free energy barrier $\Delta G$ separating the 
``two-states'' is too high to be overcome by intracellular machinery 
utilizing regular energy sources (ATP, e.g.). Thus, elucidating how the 
repeat length $\Delta l$ is related to these two properties is the main 
goal of present work. 

\bigskip
\noindent{\bf Methods}
\smallskip

\begin{figure}
\centerline{
{\epsfxsize = 6in \epsffile{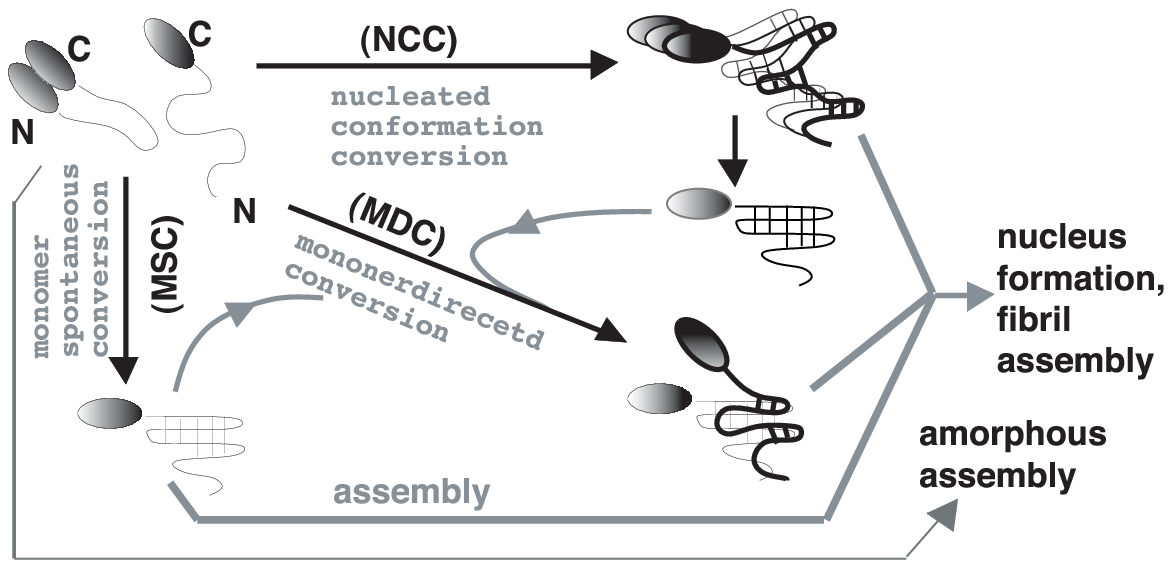}}
}\caption{}\label{1}
\end{figure}

We study the sequence-independent thermodynamics and kinetics
of an extensive homopolymer (to mimic the oligopeptide repeats) that is 
either at the tail or an inserted loop of a pathological peptide, 
Fig. \ref{1}. 
We are interested in how this homopolymer can form ordered $\beta$-sheets 
that provide the most advantageous fibril template (of course, the
homopolymer can form single-strand alpha-helix and multi-strand
amorphous aggregate, which nevertheless is less pathogenic and is not
our major concern). In particular, we identify the homopolymer length
$\Delta l$ and corresponding $\beta$-sheet topology at which the
two-state-like behavior emerges with a large free energy barrier
$\Delta G$. Furthermore, we investigate the predominant kinetics and 
patterns (Fig. \ref{1}) characterizing the transition; this knowledge can 
then be applied to therapy issues such as drug design targets and 
expected dose-response curves.

Our study proceeds in two stages. First, we study the formation of
a single-layer $\beta$-sheet (from one isolated pathological peptide)
consisting of M+1 strands each of length L residues, making use of
a model incorporating the cooperativity inherent in H-bond
formation. This cooperativity has its microscopic origin in the 
collective expelling of water molecules \cite{jose} 
from inter-chain positions in the nascent sheet \cite{jcp} that 
results in the free energy of formation of a single H-bond being 
dependent on the state of the neighboring bonds along the hairpin axis 
(intra-hairpin coupling) and between different hairpin segments along 
the ``fibril" elongation axis
(inter-hairpin coupling) \cite{jcp,PRL}. The thermodynamics is
then calculated by Monte Carlo enumeration of the partition
function \cite{PRL}. Furthermore, we utilize the idea of droplet
formation \cite{peter,droplet,droplet'} to analyze the dominant kinetic
pathway for the folding transition. Afterwards, we incorporated
sheet-sheet interactions (``stacking") and made a simple 
assumption that the free energy change due to a single H-bond formation 
within one peptide has a linear dependence on the local H-bond density
contributed by other nearby sheets. Using a mean-field approach,
we then can compute the H-bond density self-consistently. The precise 
form of our Hamiltonian as well as the details of our calculations are
available in the appendix and supplement.

\bigskip
\noindent{\bf Results and Discussion}
\smallskip

\begin{figure}
\centerline{
{\epsfxsize = 5.5in \epsffile{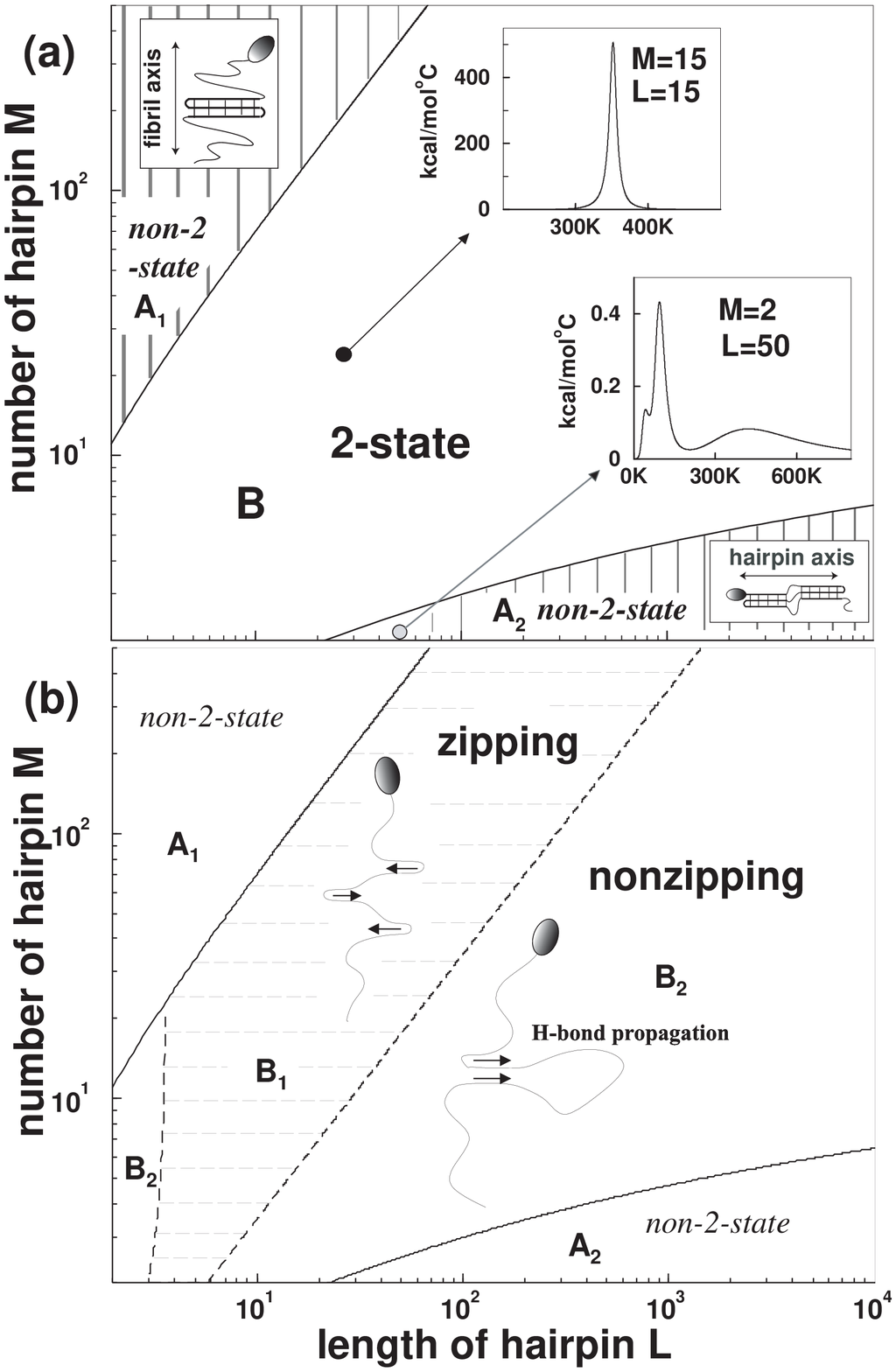}}
}\caption{}\label{2}
\end{figure}

\begin{figure}
\centerline{
{\epsfxsize = 4.5in \epsffile{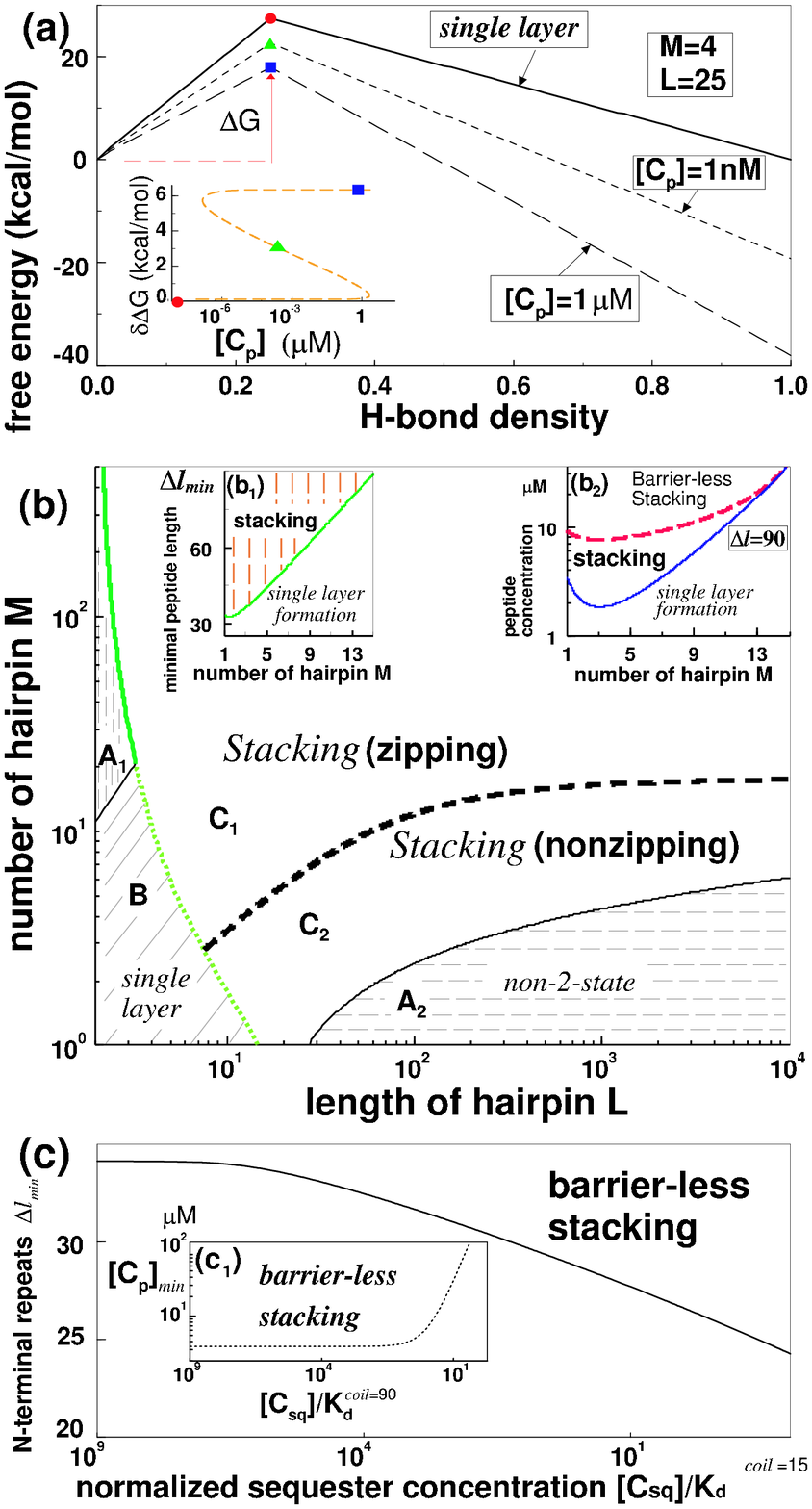}}
}\caption{}\label{3}
\end{figure}

\noindent{\bf Two-state single-layer $\beta$-sheet formation.} 
Results for this case are shown in Fig. \ref{2}. First,
there is indeed a region of two-state thermodynamics, bounded by
two regions inside which the formation of non-native
$\beta$-sheets always encounters intermediate states (and hence
can be eliminated by intracellular scavengers), Fig. \ref{2}a. This
implies, in accord with experimental observation \cite{personal},
that too long peptide N-terminal repeats do not aggregate {\it in
vivo}. Second, the kinetic barrier $\Delta G$ for two-state
transition is large ($\ge 25$ kcal/mol for number of H-bond $\ge
10^2$, Fig. \ref{3}a, e.g.), meaning that $\beta$-sheet formation in
two-state region cannot be easily reversed by intracellular
scavengers. Inside the two-state region, we also observe two
distinct patterns that dominate the kinetics, as a function of 
the topology.

\smallskip
\noindent{\bf Mechanism facilitating {\it in vivo} amyloidosis.}
The large $\Delta G$ also implies that the transition would
not occur in any reasonable time span. Thus, there must be another
mechanism to enhance non-native $\beta$-sheet formation. 
The most likely one is the inter-sheet stacking, as 
we note that folded $\beta$-sheets tend to stack into a
nearly anhydrous structure even when rich with polar side-chains
\cite{dehydrated}. This is probably because a) hydration of a
large-scale 2-dimensional structure requires extensive,
homogeneously distributed H-bond donors and acceptors to interact
with water molecules, and b) the formation of extensive
$\beta$-sheet H-bond network consumes all available backbone
electron pairs and protons (including the C$_\alpha$H$\cdots$OC
interaction \cite{jcp,CaH-OC}). Eventually, neighboring sheets 
prefer squeezing out inter-sheet water molecules. This leads to a 
synergistic H-bond formation and sheet-sheet stacking, as a function 
of peptide concentration $[C_p]$. 

\smallskip
\noindent{\bf Sheet-sheet stacking facilitate amyloidosis.}
The results in Fig. \ref{3}a show that change of $[C_p]$ can reduce
$\Delta G$ by an extent of $1/4\sim 1/2$ fold. Also, the stacking
itself has a kinetic barrier $\Delta G_s$ for a collective onset
(i.e., a phase separation between the dilute single-layer and
dense multi-layer peptide phases). This $\Delta G_s$ is different
from $\Delta G$ and can be reduced to zero if $[C_p]$ is high
enough, Fig. \ref{3}b$_2$. More importantly, in order to have stacking
effect, we found that a minimal length of peptide N-terminal
repeat $\Delta l_{min}$ is required, Fig. \ref{3}b$_1$. In other words,
the kinetics of non-native $\beta$-sheet formations for peptides
with $\Delta l <\Delta l_{min}$, is qualitatively similar to the
single-layer case. But, once $\Delta l>\Delta l_{min}$, the
kinetics can be speeded up by a synergistic inter-sheet stacking.
Consequently, non-native $\beta$-sheet formations can be more
efficient at large $\Delta l$ and high $[C_p]$ conditions. This
finding is consistent with the positive correlation between
aggregation tendency and amount of $\Delta l$, as observed both 
clinically and experimentally \cite{oligo-prion,polyQ}. 
Indeed, the minimal number of poly-glutamine repeats in Huntington 
disease is 30-40 \cite{polyQ}, in a surprisingly good agreement with 
our estimates and hence strongly suggesting that {\it in vivo} aggregation 
is driven by multi-layer $\beta$-sheet stacking. 

\smallskip
\noindent{\bf Template assembly \& Nucleated Conversion.}
The inter-sheet stacking helps $\beta$-sheet formation in high
peptide concentrations; once formed, these stacked sheets can
nevertheless dissociate if the peptide concentration is down-regulated
or fluctuating to a lower level (as a common scene at {\it in vivo}
conditions). But, with a large $\Delta G$, the dissociated sheets
can stay at stable single-layer format for a reasonable time span; 
this would allow them to serve as templates to direct other cytoplasmic
peptides into non-native $\beta$-sheet format, Fig. \ref{1}. Thus,
sheet-sheet stacking synergistically enhances non-native $\beta$-sheet
formation on both thermodynamics and kinetics prospectives.
This is a combined nucleated conformational conversion (NCC) and
mono/oligomer-directed conversion (M/O-DC) (or template assembly, TA) 
process \cite{nucleation,personal}, and might reconcile the long-standing
debate on the self-perpetuating mechanism in amyloid seed
formation \cite{nucleation}.

\smallskip
\noindent{\bf The effect of sequestering agent.}
Finally, we consider the implications of our work for the
dose-response curves of therapeutic agents. We consider the
simplest case that a scavenger can sequester unfolded mutant
peptides before they pass the transition state. In this case,
binding of the sequestering agent is {\it mutually exclusive} with
$\beta$-sheet H-bond network formation; this is presumably how
scavengers such as chaperones or polyamine
\cite{chaperone,dendrimer,chaperone-2} function. A successful
agent, then, should enhance the bottleneck (i.e., minimal mutant
peptide concentration $[C_p]_{min}$ and minimal length of repeats
$\Delta l_{min}$) for the onset of stacking. Using the mean-field
approach, we found however that at high sequester concentration,
$\Delta l_{min}$ is reduced (Fig. \ref{3}c) because of the modulation of
stacking cooperativity (whereas $[C_p]_{min}$ is increased); this
suggests that amyloid nucleation is {\it least} prohibited at
median agent concentration. Interestingly, this result seems to
correspond with the evidence that huntingtin aggregation is
eliminated only via the overexpression or deletion of chaperones
\cite{chaperone-2}.

\bigskip
\noindent{\bf Summary}

We have presented a modeling approach to the initial
seed formation in protein amyloidosis. Our work indicates that
stacking is the critical effect and that simple agents that try to
interfere with H-bond network formation may not be very effective
at preventing aggregation; alternatively, the kinetics analysis
(Fig. \ref{3}b) suggests that a potentially more effective strategy
would be to attempt to interfere with stacked non-native
$\beta$-turns or with dissociated non-native single-layer
$\beta$-sheets. Future work will take into account competition
between $\beta$-sheets and non-trivial native structures, hence
allowing for application of our ideas to a wide variety of
aggregation-prone systems. Finally, it is interesting to speculate
on the fact that even given the obvious disadvantage of an
aggregation tendency, natural selection has not replaced these
oligopeptide-repeat containing proteins by other sequence designs.
Perhaps they can serve as an essential building block for
bio-architectural construction, \cite{jcp,personal}, or an
evolutionary tool which aids in the addition of new sequence to an
existing peptide \cite{new-aa}, or a buffer to titrate chaperones
and hence to expose signaling molecules that have genetic,
structural variations buffered by chaperones to environmental
challenges \cite{genetic-buffering}.

\bigskip
\noindent{\bf Figure captions}
\smallskip
\begin{enumerate}
\item
The possible kinetics for amyloid nucleation. The
N-terminal repeats are sketched by the thin curve connecting the 
peptide (shadowed) N-, C-terminal domains. The
basic choice is whether monomers form aberrant sheets on their own
and then either form fibrils directly (monomer spontaneous
conversion) or act as templates for further monomer attachment
(monomer-directed conversion) or alternatively whether the entire
nucleus must form cooperatively (nucleated conformation
conversion). Our work suggests a modified picture in which
cooperatively formed nuclei may disassociate and act as
monomer-directed templates - see text for discussion.

\item
{\bf(a)} The phase diagram of a single-layer $\beta$-sheet, 
exhibiting the transition from two-state (region B) to non-two-state 
(A$_1$, A$_2$) behaviors as a function of sheet topology ($M$,$L$). In
region A$_1$ (A$_2$), there is a dominant intermediate ensemble
with a partially folded $\beta$-sheet droplet (unfolded bubble)
floating along the fibril (hairpin) axis. The inserts show
representative specific heat diagrams. {\bf(b)} The predominant
kinetic pathway. We computed the free energy change corresponding
to a sequence of partially-folded states differing by the addition
of one contiguous H-bond and thereby find the transition state and
folding barrier for various paths \cite{droplet,droplet'}.
Analysis shows two predominant patterns: H-bonds are initiated 1)
at several $\beta$-turns, followed by symmetric propagation
(symmetric zipping \cite{droplet}, B$_1$), or 2) at a collapsed
loop and an existing turn, followed by asymmetrically propagation
(asymmetric non-zipping, B$_2$).

\item
{\bf(a)}
The free energy profile for the predominant kinetics at a given
topology and peptide concentration $[C_p]$. $\Delta G$ is the free
energy barrier height for $\beta$-sheet formation within a single
peptide and $\delta \Delta G=\Delta G|_{[C_p]}-\Delta
G|_{[C_p]\rightarrow 0}$ (inset, dashed curve; note its
instability which leads to a sudden jump of $\delta \Delta G$
(solid curve) and hence the onset of stacking). {\bf(b)} The phase
diagram including the stacking effect. A minimal oligopeptide
repeat length $\Delta l_{min}$ is required for stacking (the thin
dotted curve and inset {\bf b$_1$}). Here $\Delta l$=(M+1)(L+2)
(assuming two C$_\alpha$ residues for one $\beta$-turn). The
kinetics is also modified by the stacking effect (symmetric
zipping: C$_1$; asymmetric non-zipping: C$_2$). {\bf(b$_2$)} The
peptide concentration $[C_p]_{stack}$ at $\Delta l$=90 required
for the onset of stacking and barrier-less stacking; the order of
$[C_p]_{stack}$ is compatible with experiment \cite{nucleation}.
{\bf(c)} The $\Delta l_{min}$ and {\bf(c$_1$)} the minimal peptide
concentration $[C_p]_{min}$ (for $\Delta l$=120) required for
barrier-less stacking as a function of agent concentration
$[C_{sq}]$. Here $K_d^{coil \rm=N}$ is the dissociation constant
between the sequestering agent and a peptide coil that can form a
totally N H-bond $\beta$-sheet.
\end{enumerate}

\newpage % figures

\newpage 

\newpage 

\hfil

%\bigskip

\newpage

\noindent{\bf Appendix}

\bigskip
\noindent{\bf The Model Hamiltonian}
\smallskip

We consider a single isolated sheet consisting of $M+1$ strands
each of length $L$ residues (i.e., number of total H-bonds
$N=ML$), with the coordinates of the $j^{\rm th}$ residue ($1\le
j\le L$) on the $i^{\rm th}$-strand ($1\le i\le M$) labeled as
$\vec{x}_{i,j}$. Since $\beta$-sheet formation primarily involves
a competition between the solvation of random coils and the
formation of a collective H-bond network, we use an {\it effective}
hydration parameter $\vec{h}_{i,j}\equiv
\vec{x}_{i+1,j}-\vec{x}_{i,j}$ to characterize the local H-bonding
between adjacent strands. Moreover, as interested in the formation of 
best $\beta$-sheet template for amyloidosis, we monitor the H-bonds between
residues labeled by the same index in the two neighboring hairpin
strands. Thus, we define a parameter $\Dij=1$ if $|h_{i,j}|\approx
0$ (i.e., no hydration) and 0 otherwise, to measure the presence
of H-bonds correctly contributing to the demanded $\beta$-sheet
structure. This leads to the following phenomenological model \cite{PRL}
\begin{eqnarray}
&&\HHb=\sum_{i,j}\Dij\left[\s f_2\Delta_{i,j\pm1} +\s
f_3\Delta_{i\pm1,j}+f_1\right]\label{Hb}\\ 
&&\Hco=\s \kappa \sum_{i,j}\Bigl[1+
\gamma\sum_{k=0}^1\Delta_{i,j-k}\Bigr]\prod_{k=\pm1} \Bigl[1
%\label{Hc}\\&&\hspace{0.5cm} 
+\gamma_1\sum_{a=0}^1\Delta_{i,j-a}
\Delta_{i+k,j}\Delta_{i+k,j-1}\Bigr]
\left|\vec{h}_{i,j}-\vec{h}_{i,j-1}\right|^2%\nonumber 
\label{Hc}
\end{eqnarray}
to describe the energy of correctly formed H-bonds and the entropy
of the unfolded coils. Here, $f_1$ determines the basic energy of
formation of a single H-bond and $\kappa$ the stiffness of an
isolated strand. The other parameters are introduced to mimic the
cooperativity inherent in H-bond network formation and, as described
elsewhere, are fitted by simulations on a more microscopic model
\cite{PRL}. In particular, $f_2,\gamma$ describe the
intra-hairpin coupling along the ``hairpin'' axis (parallel to the
stacked $\beta$-strands), and $f_3,\gamma_1$ are the inter-hairpin
coupling along the ``fibril'' axis (the fibril direction,
perpendicular to hairpin axis).

Of course, when the system becomes larger, there will be
mis-paired H-bonds generated between non-adjacent or mis-slid
adjacent strands \cite{jcp}. The lowest order correction to the
``bare'' partition function (based on Eqns.\ref{Hb},\ref{Hc}) is
due to one single mis-paired H-bond formed between any two
strands. This correction is estimated to be \cite{PRL}
\begin{eqnarray}\label{Gamma2}
\Gamma_2&\approx& \frac{(M+1)M}{2!} \times
\frac{4(2-\sqrt{2})}{(1+\gamma)^3} 
e^{-\beta f_1}L^{1/2}\sqrt{1-\Q}
%\nonumber\\ &\stackrel{\rm def}{\equiv}&
\stackrel{\rm def}{\equiv}
\frac{(M+1)M}{2!}\sigma_{\Q}
\end{eqnarray}
with $\beta=1/k_B T$, and $\Q=\sum_{ij}\Dij/N$ is the density of
correctly formed H-bonds for the sheet.

Next, to include the stacking effect, we assume that this effect
will modify the change of free energy due to the formation of a local, 
single H-bond, by an amount proportional to the density of ordered 
H-bonds on nearby sheets. This modification includes both a stacking 
energy $V_s$ arising from non-specific van der Waals interactions and 
an entropy loss $\Delta S$ due to the reduction of the radius of 
gyration between partially folded adjacent $\beta$-sheets. Details
of how we estimated the stacking effect is given in supplement file.
Overall, the Hamiltonian for the entire system follows
\begin{eqnarray}\label{H_entire}
{\rm H}&=&
\sum_s\left\{\HHb(s)+\Hco(s)\right\}
%\nonumber\\ &&\hspace{0.1cm}
+\frac{1}{2}\sum_{st}J_{st}\sum_{ij}\Dijt\left[g_1-g_2\Dijs\right]
\nonumber\\
% simplified
&\stackrel{\rm def}{\equiv}& \sum_s{\rm H}_{\text{\em
sing}}(s)+\frac{1}{2} \sum_{st}J_{st}{\rm H}_{\text{\em int}}(s,t)
\end{eqnarray}
with $J_{st}=1$ if sheets $s$, $t$ are stacked together and 0
otherwise, and $\Dijs$ is the $\Dij$ parameters for sheet $s$.
Here, we separate the entire Hamiltonian to a
single-sheet part ${\rm H}_{\text{\em sing}}$ and an inter-sheet
coupling part ${\rm H}_{\text{\em int}}$.
Also, for future convenience, we have defined notations
$g_1=\frac{k_B T}{N}\ln\left[\frac{\Omega_0}{\Omega_N}\right]$ and
$g_2=N^{-1}[-V_s+g_1]$, where $\Omega_0$ and $\Omega_N$ are the
volume of completely unfolded coil and folded $\beta$-sheet, respectively.

\bigskip
\noindent{\bf The Partition Function}
\smallskip

\noindent{\bf Without Stacking Effect.}
Given the above expression, we transform the $\beta$-sheet into a
system with many coupled hairpins. Then, for a given choice of
topology $(M,L)$ and contact configuration $\D$ (which then yields
a particular H-bond density $\Q$), we found that the partition
function can be dissected in terms of three parts
\begin{eqnarray}\label{Z1}
&&\Z = Z_{\rm Hb}Z_{coil}\left[1+\Gamma_2\right]
\end{eqnarray}
that allows us to compute contributions from the entropy of coiled
part (in $Z_{coil}$), the energy of partially formed H-bond
(in $Z_{\rm Hb}$), and the H-bond mis-pairing effect (in
$\Gamma_2$). Again, the details of derivation can be found in
supplement file.

For the thermodynamics, we are interested in whether at the
folding temperature $T_f$ (at which the partition function $\zq$
of a random coil, $\Q=0$, and the completely folded state, $\Q=1$,
are equally weighted), there are other competing intermediates
$Z_\Qi$. In principle, for a two-dimensional $\beta$-sheet
structure, there are two possible competing intermediates with a
partially folded droplet (unfolded bubble) floating along the
fibril (hairpin) axis with translational entropy large enough to
compete with the completely un/folded states $Z_0$, $Z_1$. The
phase boundaries separating the dominance of these ensemble are
illustrated in Fig. \ref{2} with representative specific heat diagrams
inserted. Our approach to the study of the kinetics of folding is
described in the caption to Fig. \ref{2} and elsewhere
\cite{droplet,droplet'}.

\smallskip
\noindent{\bf With Stacking Effect.}
Using variational mean field theory to approach the thermodynamics, 
we obtained the following self-consistent equation
\begin{eqnarray}
\q&=&\frac{e^{\beta\mu}\sum_\Q \Q\zqq}{1+e^{\beta\mu}\sum_\Q \zqq}
\label{mean-field}\\ 
\zqq&=&Z_\Q[0]
e^{-N\left[2\beta V_s\Q 
   +N^{-1}(1-2\Q)\ln\frac{\Omega_0}{\Omega_N} \right]\q}
%\nonumber\\ &\stackrel{\rm def}{\equiv}& 
\stackrel{\rm def}{\equiv}
Z_\Q[0]e^{h_1\q+h_2\Q\q}\label{Z2}
\end{eqnarray}
where $Z_\Q[0]$ is the bare partition function without the
stacking effect and $\mu$ is the peptide chemical potential. 
Apparently, $\mu$ is related to the peptide concentration $[C_p]$ 
and the relation is estimated as
\begin{eqnarray}\label{Cp1}
[C_p]\Omega_0 &\approx&
\frac{e^{\beta\mu}Z_0[\Q]}{1+e^{\beta\mu}Z_0[\Q]}
\biggr|_{\q\approx 0}
\end{eqnarray}
To identify where the two-state-like behavior emerges, we examined
if any of the aforementioned intermediates dominates at the
temperature $T_f$ where $Z_0[0]=Z_1[0]$. Apparently, if $\Qi$ is
the H-bond density of the dominant ensemble, we would have
$\q=\Qi$. This implies that the two-state behavior exists if
$\forall \Qi$, we have ($0<\Qi<1$)
\begin{eqnarray}\label{Z3}
Z_\Qi[\Qi]&=&Z_\Qi[0]e^{(h_1+h_2\Qi)\Qi}
%\nonumber\\ &<&
< 
{\rm max}\left[Z_0[\Qi],Z_1[\Qi]\right]
\end{eqnarray}
for a given topology $(M,L)$ and temperature $T_f$. Solving this
requirement, we obtained the phase diagram Fig. \ref{3}a; likewise, the
predominant kinetic patterns can be computed in the 2-state
region (see supplement for details).

\bigskip
\noindent{\bf Estimating the Minimal Length for Stacking}
\smallskip

At the two-state region, the ensemble sum in 
(\ref{mean-field}) can be reduced as $\sum_\Q\zqq\approx
Z_0[\q]+Z_1[\q]$; thus after re-arrangement, Eqn.
(\ref{mean-field},\ref{Z2}) can be re-formulated into
\begin{eqnarray}
[C_p]\Omega_0&=&\frac{\q}{1-\q}e^{-f_{stack}\q}\label{mu-Q}
\end{eqnarray}
where $f_{stack}=2[N\beta V_s+\ln(\Omega_0/\Omega_N)]$.
Then, from a previous study \cite{self-consistent,self-consistent'},
we realize that Eqn. (\ref{mu-Q}) can yield a ``van-der-Waals''
loop in the $([C_p],\q)$ diagram and hence a phase separation
(coexistence) effect. The phase separation referred to here is a
separation between the dilute random coiled phase and the dense
stacking phase. Numerical calculation showed that this requires
a minimal length $\Delta l_{min}$, as shown in Fig. \ref{3}a$_1$. 
The peptide concentration that allows for phase separation can
also be computed (see supplement for details); this
yields Fig. \ref{3}a$_2$.

\bigskip
\noindent{\bf Effect of Sequesterers.}
\smallskip

The effect of sequesterers on the $\beta$-sheet partition function
is estimated as the following
\begin{eqnarray}\label{Z4}
\zqq&\rightarrow& \zqq\left[1+
\frac{(1-\Qx)[C_{sq}]}{K_d\left[\{\Delta_{ij}\}\right]} \right]
\end{eqnarray}
where $[C_{sq}]$ is the sequesterer concentration,
$K_d\left[\{\Delta_{ij}\}\right]$ is the $\beta$-sheet
configuration-dependent dissociation constant with the sequesterer, 
$\Qx$ is the averaged H-bond density from other sheets nearby the
one interacting with sequesterers. Here the prefactor
$(1-\Qx)$ indicates that binding of sequesterer is unlikely to
occur if the target peptide is tightly surrounded by other sheets 
(as they tend to stack together into an anhydrous, dense aggregate). 

\smallskip
\noindent{\bf Mutually exclusive effects.} One sequesterer might
generally bind to multiple amide (or carbonyl) groups on a single targeted 
peptide; here for simplicity, we assume that the binding of each 
individual sequesterer functional site to the peptide amide (or carbonyl) 
group occurs in an independent manner. Moreover, we assume that the 
binding is {\it mutually exclusive} with local H-bond formation, as
presumably how scavengers such as polyamine \cite{dendrimer}
function. Then, for a pathological peptide that can form N H-bonds 
in total, assuming $K_d^{coil}$ as the dissociation constant for a 
fully coiled peptide, we found that the peptide concentration for
sheet-sheet stacking is modified (see supplement for detailed
derivation)
\begin{eqnarray}
&&[C_p]\Omega_0 = \left[1+\frac{[C_{sq}]}{K_d^{coil}
}\left(1-\q\right)\right] 
%\nonumber\\ &&\hspace{1.0cm}\times
\frac{\q}{1-\q}e^{-\q f_{stack}} \label{Cp3}
\end{eqnarray}

Indeed, we found that the cooperativity in sheet-sheet stacking
is enhanced by the presence of sequesterers. This is because the
binding of sequesterer and $\beta$-sheet H-bond formation as well
as stacking are {\it mutually exclusive}. Thus, the thermodynamic 
weighting of intermediates with partially formed H-bonds and weakly 
stacked $\beta$-sheets are significantly suppressed; only completely
folded and well-stacked sheets can escape from the attack of
sequesterers.
In the presence of sequesterer, we are interested in its
dose-response curve. Specifically, we compute the minimal peptide
concentration $[C_p]_{min}$ as well as minimal length of repeats
$\Delta l_{min}$ required for the downhill stacking. The numerical
results are shown in Fig. \ref{3}b.

\end{document}

\def\ra{{r_{\rm Hb}^c}}
\def\rb{{r_{\rm Hb}^*}}
\def\rc{{r_{\rm Hb}^{sol}}}
\def\rx{{r_{\rm Hb}^x}}
\def\rhb{r_{\rm Hb}}
\def\s{{\frac{1}{2}}}
\def\x{{x_{\rm Hb}}}
\def\Ra{{r_{\rm sc}^c}}
\def\Rb{{r_{\rm sc}^*}}
\def\Rc{{r_{\rm sc}^{sol}}}
\def\Rx{{r_{\rm sc}^x}}
\def\X{{x_{\rm chiral}}}
\def\dt{{\Delta t}}
\def\go{{G$\rm\bar{o}$}}
\def\LJf{{LJ$^{fix}$}}
\def\LJa{{LJ$^{ang}$}}
\def\Solf{{Sol$^{fix}$}}
\def\Sola{{Sol$^{ang}$}}
\def\Qi{{Q_{int}}}
\def\qi{{\frac{Q_{int}}{N}}}
\def\q{{\langle\bar{Q}\rangle}}
\def\zqq{{Z_Q\left[\langle\bar{Q}\rangle\right]}}
\def\pdd{{p_{dd}}}
\def\cdd{{[C_{dd}]}}
\def\ps{{p_{s}}}
\def\cs{{[C_{s}]}}
\def\NHOC{{NH$\cdots$OC }}
\def\vHb{{v_{\rm Hb}}}
\def\fsq{{f_{i,j}^{\text{\em sq}}}}
\def\psq{{P_{\text{\em sq}}}}
\def\HVs{{\rm{H}_{V_s}}}
\def\HDS{{\rm{H}_{\Delta S}}}
\def\ra{{r_{\rm Hb}^c}}
\def\rb{{r_{\rm Hb}^*}}
\def\rc{{r_{\rm Hb}^{\text{\em sol}}}}
\def\rx{{r_{\rm Hb}^x}}
\def\rhb{r_{\rm Hb}}
\def\s{{\frac{1}{2}}}
\def\x{{x_{\rm Hb}}}
\def\Ra{{r_{\rm sc}^c}}
\def\Rb{{r_{\rm sc}^*}}
\def\Rc{{r_{\rm sc}^{\text{\em sol}}}}
\def\Rx{{r_{\rm sc}^x}}
\def\X{{x_{\rm chiral}}}
\def\dt{{\Delta t}}
\def\go{{G$\rm\bar{o}$}}
\def\LJf{{LJ$^{fix}$}}
\def\LJa{{LJ$^{ang}$}}
\def\Solf{{Sol$^{fix}$}}
\def\Sola{{Sol$^{ang}$}}
\def\Qi{{\bar{Q}_{\text{\em int}}}}
\def\qi{{\frac{Q_{\text{\em int}}}{N}}}
\def\Q{{\bar{Q}}}
\def\Qx{{\bar{Q'}}}
\def\Qij{{{Q'}_{ij}}}
\def\Qijk{{{Q'}_{ij,(k)}}}
\def\q{{\langle\bar{Q}\rangle}}
\def\HHb{{\rm H_{{\rm Hb}}}}
\def\Hco{{{\rm H}_{\text{\em coil}}}}
\def\UHb{{{\rm U_{Hb}}(r,\x)}}
\def\Dij{{\Delta_{i,j}}}
\def\Dijs{{{\Delta}_{ij,(s)}}}
\def\Dijt{{{\Delta}_{ij,(t)}}}
\def\D{{\{\Delta_{i,j}\}}}
\def\Z{{Z\left[\{\Delta_{i,j}\}\right]}}
\def\zqq{{Z_{\bar{Q}}\left[\langle\bar{Q}\rangle\right]}}
\def\zq{{Z_{\bar{Q}}}}
\def\M{{M^*}}
\def\sx{{\sigma_{\frac{1}{L}}}}
\def\pdd{{p_{dd}}}
\def\cdd{{[C_{dd}]}}
\def\ps{{p_{s}}}
\def\cs{{[C_{s}]}}
\def\NHOC{{NH$\cdots$OC }}
\def\vHb{{v_{\rm Hb}}}
\def\fsq{{f_{i,j}^{\text{\em sq}}}}
\def\psq{{P_{\text{\em sq}}}}
\def\HVs{{\rm{H}_{V_s}}}
\def\HDS{{\rm{H}_{\Delta S}}}

\title{Supporting information for ``Non-native $\beta$-sheet formation: 
insights into protein amyloidosis''}

\author{Chinlin Guo$^{\dag\ddag}$, Herbert Levine$^\ddag$, 
and David A. Kessler$^*$}
\affiliation{ $^\dag$Department of Molecular Cell Biology, 
Harvard University,
16 Divinity Avenue, Room 3007, Cambridge, MA 02138} 
\affiliation{ $^\ddag$Department of Physics, University of
California, San Diego
9500 Gilman Drive, La Jolla, CA 92093-0319} 
\affiliation{ $^*$Department of Physics, Bar-Ilan
University, Ramat-Gan, Israel}

\maketitle
%\centerline{\bf Supporting Information}

\section{Estimating the Stacking Effect}\label{A1}

To proceed, we first notice that $V_s$ is actually
sequence-dependent. In general, energies of this type range from
$-10^{-2}$ to $-0.2$ kcal/mol \cite{Fersht-sup}. As we are interested
in the generic behavior independent of sequence, we take $V_s =
-0.1$ kcal/mol as the size of the non-specific
C$_\alpha$-C$_\alpha$ coupling \cite{Fersht-sup} to mimic the
stacking energy. Since this non-specific interaction between two
H-bond units (from two different, nearby sheets) occurs only when
both units have formed H-bonds, it contributes an energy between
two stacked sheet $s$ and $t$: $\HVs(s,t)=\sum_{i,j}\Dijs \Dijt
V_s$, where $\Dijs$ ($\Dijt$) is for the $(i,j)^{\rm th}$ H-bond
at sheet $s$ ($t$). The stacking energy for the entire system then
follows $E_{V_s}=\frac{1}{2}\sum_{st}J_{st}\HVs(s,t)$ where
$J_{st}=1$ if $s$, $t$ are stacked sheets and 0 otherwise. Note
that there are at most two sheet $t$'s that can stack with sheet
$s$.

Second, we notice that the entropy reduction in stacked sheets
arises merely from their structural conflicts. When the sheets are
all in their completely folded state, there is no entropy
reduction upon stacking since they are already in their minimal
entropy state. Similarly, wandering of random coils is less
prohibited when contiguous to a completely unfolded sheet,
compared to a partially or completely folded one. For simplicity,
we estimate the entropy of different conformations based on their
volume. As an example, the volume of a fully unfolded coil,
$\Omega_0$, can be estimated from the random-flight chain model
\cite{protein-sup} $\Omega_0\approx[\sqrt{l}R_{\rm C_\alpha}]^3$
where $l$ is the number of total residues at the mutant peptide
N-terminal repeats, and the unit length for radius of gyration
$R_{{\text C}_\alpha}\approx 11.4{\text \AA}$ \cite{protein-sup}.
Likewise, the volume $\Omega_N$ of a completely folded
$\beta$-sheet with a total of $N$ H-bonds ($N=ML$) can be
estimated as $\Omega_N=N\vHb$ where $\vHb\approx 4.8\times
3.8\times 10.0{\text \AA}^3$ is the unit volume of a single H-bond
measured in a densely-packed $\beta$-sheet aggregate
\cite{dehydrated-sup}.

Now, to estimate the entropy reduction and keep the formula
simple, we use a global term instead of counting all local effects
as done in the case of $V_s$. Specifically, for a given sheet $s$
that is surrounded by two other sheets, indexed by  $t$, we
introduce the quantity $\Qx=\sum_{t}\sum_{ij}\Dijt/(2N)$ to
account for the averaged H-bond density in these two sheets, as a
global indication of how well they form $\beta$-sheet structures.
Moreover, for simplicity we assume that the entropy loss of sheet
$s$ can be roughly accounted for by a progressive volume reduction
from $\Omega_0$ to $\Omega_N$ due to the structural conflict with
its stacking neighbors, and for which we take the simple form
$\HDS(s) = (1-\Q)\Qx k_BT\ln[\Omega_0/\Omega_N]$. Here, the factor
$(1-\Q)\Qx$ indicate the structural conflict between sheet $s$ and
its contiguous neighbors. Of course, one can devise a more
complicated formula to account for how the presence of H-bonds
affects the entropy reduction, but this simple form will suffice
for our present needs. 

Overall, the Hamiltonian for the entire
system follows
\begin{eqnarray}\label{H_entire-sup}
{\rm H}&=&\sum_s\left[\HHb(s)+\Hco(s)\right]+E_{V_s}
%\nonumber\\&&\hspace{0.1cm} 
+\frac{k_B T}{2N}\ln\left[\frac{\Omega_0}{\Omega_N}\right]
\sum_{st}J_{st}\Dijt\left[1-\frac{\Dijs}{N}\right] \nonumber\\
&\stackrel{\rm def}{\equiv}&
\sum_s\left\{\HHb(s)+\Hco(s)\right\}
%\nonumber\\ &&\hspace{0.1cm}
+\frac{1}{2}\sum_{st}J_{st}\sum_{ij}\Dijt\left[g_1-g_2\Dijs\right]\nonumber\\
&\stackrel{\rm def}{\equiv}& \sum_s{\rm H}_{\text{\em
sing}}(s)+\frac{1}{2} \sum_{st}J_{st}{\rm H}_{\text{\em int}}(s,t)
\end{eqnarray}
with the notation defined in main text.

\section{Estimating the Partition Function}\label{A2}

\subsection{Without Stacking Effect}\label{A2-1}

Given the Hamiltonian 
\begin{eqnarray}
&&\HHb=\sum_{i,j}\Dij\left[\s f_2\Delta_{i,j\pm1} +\s
f_3\Delta_{i\pm1,j}+f_1\right]\label{Hb-sup}\\ &&\Hco=\s \kappa
\sum_{i,j}\Bigl[1+
\gamma\sum_{k=0}^1\Delta_{i,j-k}\Bigr]\prod_{k=\pm1} \Bigl[1
%\label{Hc-sup}\\&&\hspace{0.5cm} 
+\gamma_1\sum_{a=0}^1\Delta_{i,j-a}
\Delta_{i+k,j}\Delta_{i+k,j-1}\Bigr]
\left|\vec{h}_{i,j}-\vec{h}_{i,j-1}\right|^2,
%\nonumber
\label{Hc-sup}
\end{eqnarray}
we can transform the $\beta$-sheet into a
system with many coupled hairpins. Then, for a given choice of
topology $(M,L)$ and contact configuration $\D$ (which then yields
a particular H-bond density $\Q$), we further dissected each of
these hairpins into one or several unfolded coils (``bubbles'')
separated by successively folded segments (``droplets''). The
partition function for these unfolded coils is estimated by
previously developed methods \cite{droplet-sup,droplet'-sup}.
Specifically, we define a functional $M_i(j_1,j_2)$ as the
Gaussian path integral (based on Eqn.\ref{Hc-sup}) for a coil running
from residue index $(i,j_1+1)$ to $(i,j_2-1)$ with both $(i,j_1)$,
$(i,j_2)$ being H-bonded and a functional $W_i(j_1,j_2)$ for a
continuously H-bonded segment from index $(i,j_1)$ to $(i,j_2)$
(based on Eqn.\ref{Hb-sup})
\begin{eqnarray}
M_i(j_1,j_2)&=&\prod_{j=j_1+1}^{j_2-1} \int
\left[{\beta\kappa\over2\pi}\right]^{3\over2}
d\vec{h}_{i,j}(1-\Dij) \nonumber\\&&\hspace{2.0cm}\times e^{-\beta
\HHb(\D)}\\ W_i(j_1,j_2)&=& \prod_{j=j_1}^{j_2} \int
\left[{\beta\kappa\over2\pi}\right]^{3\over2} d\vec{h}_{i,j}\Dij
e^{-\beta \Hco(\D)}\label{W-sup}
\end{eqnarray}

These Gaussian integrals can be worked out as shown in Ref.
\cite{PRL-sup,droplet-sup,droplet'-sup}. The partition function can
be broken up into successive multiplication of these functionals
\begin{eqnarray}\label{Z1-sup}
&&\Z = \prod_{i,j}\int
\left[{\beta\kappa\over2\pi}\right]^{3\over2} d\vec{h}_{i,j}
\left[(1-\Dij)+\Dij\right]
%\nonumber\\ &&\hspace{0.5cm} \times
e^{-\beta
[\HHb(\D)+\Hco(\D)]}\left[1+\Gamma_2(\Q)\right]\nonumber\\ &&=
\left\{\prod_i \left[W_i(0,j_1) M_i(j_1,j_2)
W_i(j_2,j_3)\cdots\right] \right\} 
%\nonumber\\ &&\hspace{1.0cm} \times
\left[1+\Gamma_2(\Q)\right] \stackrel{\rm def}{\equiv}
Z_{\rm Hb}Z_{coil}\left[1+\Gamma_2\right]
\end{eqnarray}
where $(i,j_1), (i,j_2), \dots$ are the end points for the
successive folded segments. Thus, in this way, we can dissect the
free energy in terms of contributions from the Gaussian integral
(in $Z_{coil}$), from the {\it Ising-model-like} energy component
(in $Z_{\rm Hb}$), and from the H-bond mis-pairing effect (in
$\Gamma_2$). This allows us to fit the parameters with simulations
on a microscopic model \cite{PRL-sup}; once obtained, these
parameters can be used to retrieve the entire partition function
and hence the density of states using Multicanonical Monte Carlo
sampling \cite{MC-sup}.

\subsection{With Stacking Effect}\label{A2-2}

In the presence of stacking effect, the partition functional for
the entire system reads
\begin{eqnarray}\label{Zt1-sup}
{\cal Z}&=&\int{\cal D}h e^{-\beta\sum_s\left[ {\rm H}_{\text{\em
sing}}(s)+\frac{1}{2} \sum_tJ_{st}{\rm H}_{\text{\em
int}}(s,t)\right]}
\end{eqnarray}
where $\int{\cal D}h$ indicates integration for the entire
$\vec{h}_{i,j}(s)$ vector space. To compute the thermodynamics, we
use a Gaussian trick implemented in Ref.
\cite{self-consistent-sup,self-consistent'-sup} to decouple one of the
inter-sheet terms. Briefly, we insert a Gaussian integral identify
in Eqn.(\ref{Zt1-sup}) to separate the $\Dijs\Dijt$ terms
\begin{eqnarray}\label{Zt2-sup}
{\cal Z}&\rightarrow&\int{\cal D}h e^{-\beta\sum_s\left[{\rm
H}_{\text{\em sing}}(s) +\frac{g_1}{2}\sum_t J_{st}\Dijt\right]}
%\nonumber\\ &&\hspace{0.1cm}\times 
\int{\cal D}\zeta
e^{-\frac{1}{2\beta g_2}\sum_{st}J_{st}^{-1}\sum_{ij}
\zeta_{ij,(s)}\zeta_{ij,(t)}}
\nonumber\\ &&\hspace{1.0cm}\times
e^{\sum_s\Dijs\zeta_{ij,(s)}}
\end{eqnarray}
Then, using a transform $\zeta_{ij,(s)}=\beta
g_2\sum_tJ_{st}\eta_{ij,(t)}$ and the identity $\sum_{t}
J_{st}\Dijt=2N\Qx(s)$, we have
\begin{eqnarray}\label{Zt3-sup}
{\cal Z}&\rightarrow& \int{\cal D}\eta e^{-\frac{\beta g_2}{2}\sum_{st}
J_{st}\sum_{ij}\eta_{ij,(s)}\eta_{ij,(t)}}
%\nonumber\\&&\hspace{0.1cm}\times 
\int{\cal D}h e^{-\beta\sum_s\left[{\rm
H}_{\text{\em sing}}(s)+Ng_1\Qx(s)\right]} 
\nonumber\\ &&\hspace{1.0cm}\times 
e^{\beta g_2\sum_{st}J_{st}\sum_{ij}\Dijs\eta_{ij,(t)}}
\end{eqnarray}

Next, we use variational mean field theory to approach the
thermodynamics. First, we set $\Qx=\q$ in Eqn.(\ref{Zt3-sup}) as the
thermodynamically averaged H-bond density. Second, we rewrite the
partition function into a form of grand canonical ensemble by
inserting a chemical potential $\mu$ for each peptide. These steps
allow us to decouple the partition function Eqn.(\ref{Zt3-sup}) into a
product of independent single peptide partition function where
each peptide interacts with a background field $\{\eta\}$.
Finally, the connection between $\{\eta\}$ and $\q$ can be found
by a variational approach on $\{\eta\}$ similar to the procedure
used in Ref. \cite{self-consistent-sup,self-consistent'-sup}. This
yields a self-consistent relation $\eta_{ij,(s)}=\langle
\Dijs\rangle$ where $\langle \Dijs\rangle$ is the thermodynamic
expectation value, which by definition is equivalent to $\q$ in
the sense of mean-field approach. Overall, one can work out the
following self-consistent equation
\begin{eqnarray}
\q&=&\frac{e^{\beta\mu}\sum_\Q \Q\zqq}{1+e^{\beta\mu}\sum_\Q \zqq}
\label{mean-field-sup}\\ \zqq&=&
Z_\Q[0]e^{\beta\left[-g_1+2g_2\sum_{ij}\Dij\right]\q} \nonumber\\
&=&Z_\Q[0]e^{-N\left[2\beta V_s\Q
+N^{-1}(1-2\Q)\ln\frac{\Omega_0}{\Omega_N} \right]\q}\nonumber\\
&\stackrel{\rm def}{\equiv}& Z_\Q[0]e^{h_1\q+h_2\Q\q}\label{Z2-sup}
\end{eqnarray}
where $Z_\Q[0]$ is defined as in the main text.
Note that the entropy effect is scaled by $N^{-1}$ and hence is less 
important than the energy term.

The loss of translational entropy of one peptide due to the
presences of other peptides is absorbed into the chemical
potential $\mu$ in this mean-field approach
\cite{self-consistent-sup}. To estimate the relation between
$\mu$ and peptide concentration $[C_p]$, we use a dilute
gas of weakly folded peptides as the basic state which then
``competes'' with aggregate formation. In other words, the peptide
concentration is that before the onset of aggregation.
Specifically, we estimated the probability of finding a partially
folded peptide in a unit volume that can be caught by other
peptides. The size of this volume is estimated roughly to
$V_\Q\approx 4\pi [R_G(\Q)]^3/3$ with $R_G(\Q)$ and $\Q$ denoting
the radius of gyration and H-bond density of that peptide,
respectively. $R_G(\Q)$ is further estimated by a linear
interpolation $R_G(\Q)=(1-\Q)\sqrt{l}R_{\rm
C_\alpha}+\Q[\vHb]^{1/3}$ between the volume estimates of fully
un/folded $\beta$-sheets. Then, $[C_p]$ is connected to $\mu$ via
the probability $P_p$ of finding a peptide in volume
$V_{\Q\approx0}$
\begin{eqnarray}\label{Cp1-sup}
[C_p]V_{\Q\approx0}
&\approx&\frac{1}{\beta}\frac{\partial}{\partial\mu}
\ln\left[1+e^{\beta\mu}\sum_{\Q\ll 1}Z_\Q[\q]\right]_{\q\approx
0}\nonumber\\ &\approx&
\frac{e^{\beta\mu}Z_0[\Q]}{1+e^{\beta\mu}Z_0[\Q]}
\biggr|_{\q\approx 0}
\end{eqnarray}
where $V_{\Q\approx0}\approx \Omega_0$. Thus, for any given M, L, and T, 
we can compute $Z_\Q[0]$ by Monte Carlo simulation, and then combine 
the given $[C_p]$ to solve $\q$ self-consistently.

\section{Minimal Length for Stacking}\label{A3}

At the two-state region, the ensemble sum in Eqn.
(\ref{mean-field-sup}) can be reduced as $\sum_\Q\zqq\approx
Z_0[\q]+Z_1[\q]$; thus after re-arrangement, Eqn.
(\ref{mean-field-sup},\ref{Z2-sup}) can be re-formulated by defining an
effective chemical potential $\mu'$
\begin{eqnarray}
e^{\beta\mu'}&\stackrel{\rm def}{\equiv}&
\frac{Z_1[0]e^{\beta\mu-\q\ln\frac{\Omega_0}{\Omega_N}}}
{1+Z_0[0]e^{\beta\mu-\q\ln\frac{\Omega_0}{\Omega_N}}}\label{mu1-sup}\\
\q&=& \frac{e^{\beta\mu}Z_1[\q]}
{1+e^{\beta\mu}\left(Z_0[\q]+Z_1[\q]\right)}\nonumber\\
&=&\frac{e^{\beta \mu'}e^{ 2\left [N\beta V_s
+\ln\frac{\Omega_0}{\Omega_N}\right]\q}} {1+e^{\beta \mu'}e^{
2\left [N\beta V_s +\ln\frac{\Omega_0}{\Omega_N}\right]\q}}
\nonumber\\ &\stackrel{\rm def}{\equiv}& \frac{e^{\beta
\mu'+f_{stack}\q}}{1+e^{\beta \mu'+f_{stack}\q}}
\label{mean-field-2-sup}\\ \Rightarrow \beta\mu'&=&-f_{stack}\q
-\ln\left[\frac{1}{\q}-1\right] \label{mu-Q-sup}
\end{eqnarray}
where we have used the condition that the temperature is set at
$Z_0[0]=Z_1[0]$. Then, from Eqn. (\ref{Cp1-sup},\ref{mu1-sup}) we have
\begin{eqnarray}\label{Cp2-sup}
[C_p]V_{\Q\approx0} &\approx&
\frac{e^{\beta\mu}Z_0[\Q]}{1+e^{\beta\mu}Z_0[\Q]}
\biggr|_{\q\approx 0}\nonumber\\
&=&\frac{Z_0[0]e^{\beta\mu-\q\ln\frac{\Omega_0}{\Omega_N}}}
{1+Z_0[0]e^{\beta\mu-\q\ln\frac{\Omega_0}{\Omega_N}}} \equiv
e^{\beta\mu'}
\end{eqnarray}

After a straightforward calculation, one can show that the van-der-Waals 
in $(\mu',\q)$ diagram occurs only if $f_{stack}> 4$. Numerically,
one can show that $f_{stack}$ has a dependence on the
$\beta$-sheet topology as well as the mutant polymer length
$\Delta l$ (the N-terminal oligopeptide repeats that construct the
non-native $\beta$-sheet). Thus, the criteria $f_{stack}> 4$
yields a minimal length $\Delta l_{min}$ requirement for the onset
of stacking effect, as shown in Fig. 3a$_1$. Also, from the
abovementioned study \cite{self-consistent-sup,self-consistent'-sup}, we
note that Eqn. (\ref{mu-Q-sup}) has one spinodal point at
$\q_{s-}=[1-\sqrt{1-4/f_{stack}}]/2$ where the barrier for the
onset of stacking is zero (downhill stacking), and one binodal
point $\q_{b-}=[1-\xi]/2$ with $2\xi=\tanh[f_{stack}\xi/2]$ where
the dilute and dense phases are equally weighted. The
corresponding peptide concentrations $[C_p]$ for these two points
can be computed from Eqns. (\ref{mu1-sup},\ref{mu-Q-sup},\ref{Cp2-sup})
\begin{eqnarray}\label{Cp-sup}
[C_p]V_\q&=&\frac{\q}{1-\q}e^{-\q f_{stack}} \ ,
\end{eqnarray}
yielding Fig. 3a$_2$.

\section{Effect of Sequesterers}\label{A4}

To model the effect of sequesterers, we note that the partition
function for one protein A when interacting with another one B,
$A+B\leftrightarrow AB$, has a general form
$Z_A=Z_A[0]+\frac{[C_B]}{K_d}Z_A[1]$ where $Z_A[0]$, $Z_A[1]$ are
the partition function of protein A in the absence, presence of
protein B, respectively, and $K_d$ is the dissociation constant of
the reaction. More generally, the second term can be expressed as
the sum in which various $K_d$'s, each representing a different
configuration of protein A, are included. Then, examining the
effect of one sequesterer interacting with one non-native
$\beta$-sheet, we can see that it modifies the partition function
eqn.(\ref{Z1-sup}) as follows:
\begin{eqnarray}\label{Z4-sup}
\zqq&\rightarrow& \zqq\left[1+
\frac{(1-\Qx)[C_{sq}]}{K_d\left[\{\Delta_{ij}\}\right]} \right]
\end{eqnarray}

To express the form of $K_d\left[\{\Delta_{ij}\}\right]$, we
further assume that the binding of each individual sequesterer
functional site to the peptide amide (or carbonyl) group occurs in
an independent manner. This leads to a multiplicative expansion of
the second term in eqn.(\ref{Z4-sup})
\begin{eqnarray}\label{Kd-sup}
\frac{[C_{sq}]}{K_d\left[\{\Delta_{ij}\}\right]} =e^{\beta
\mu_{sq}}\left\{ \prod_{i,j}\left[1-\Delta_{ij}\right]
\left[1+e^{\beta V_{sq}}\right]-1\right\}
\end{eqnarray}
Here, $\mu_{sq}$ is the sequesterer chemical potential. The factor
$[1-\Delta_{ij}]$ indicates that a local interaction between
sequesterer and the peptide occurs only if the local H-bond is not
formed, which then allows a local binding or unbinding via a
binding energy $V_{sq}$ roughly estimated to be $-2.8$ kcal/mol
comparable with the order of the H-bond energy (which is
presumably how sequesterer binds to the amide or carbonyl group).
Finally, the term ``-1'' accounts for the complete non-binding
event and ensures at least one binding between the sequesterer and
the peptide. In general, one can include more complicated binding
pattern but here we restrict our modeling to this simplest case.

Then, for a mutant peptide that can form N H-bonds in total,
we can rewrite eqn.(\ref{mu1-sup},\ref{mu-Q-sup},\ref{Cp-sup}) as
\begin{eqnarray}
&&\beta\mu'= -f_{stack}\q
+\ln\left[1+\frac{[C_{sq}]}{K_d^{coil}}\left(1-\q\right)\right]
%\nonumber\\&&\hspace{1.0cm} 
-\ln\left[\frac{1}{\q}-1\right]
\label{mu2-sup}\\ 
&&[C_p]V_\q = \left[1+\frac{[C_{sq}]}{K_d^{coil}
}\left(1-\q\right)\right] 
%\nonumber\\ &&\hspace{1.0cm}\times
\frac{\q}{1-\q}e^{-\q f_{stack}} \label{Cp3-sup}
\end{eqnarray}